\documentclass[twocolumn,
preprintnumbers,amsmath,amssymb
]{revtex4}
\usepackage{graphicx}

\usepackage{xcolor}

\begin{document}
\title{Fractional Langevin equation from damped bath dynamics
}
\author{Alex V. Plyukhin} 
\email{aplyukhin@anselm.edu} 
\affiliation{
Saint Anselm College, Manchester, New Hampshire} 
\date{\today}

\begin{abstract}

We consider the stochastic dynamics of a system linearly coupled to a hierarchical thermal bath with
two well-separated inherent timescales: one slow, and one fast.
The slow part of the bath is modeled as a set of harmonic oscillators and taken into account explicitly, while the effects of the fast part of the bath are simulated by dissipative and stochastic Langevin forces, uncorrelated in space and time, acting on oscillators of the slow part of the bath.
We demonstrate for this model the robust emergence of a fractional Langevin equation with a power-law decaying memory kernel. The conditions of
such an emergence and the specific value of
the fractional exponent 
depend only
on the asymptotic low-frequency spectral properties of the slow part of the bath.


\end{abstract}


\maketitle

The fractional Langevin equation (FLE) is a special, and important, type of the generalized (non-Markovian) Langevin equation~\cite{Zwanzig} for a dynamical variable $A$ coupled to a thermal bath,
\begin{eqnarray}
\dot A(t)=-\int_0^t K(t-\tau) \,A(\tau)\,d\tau +F(t),
\label{GLE}
\end{eqnarray}
with the algebraically decaying memory kernel $K(t)$. 
We shall focus on kernels with 
the decay exponent $\alpha$ 
being less than one,
\begin{eqnarray}
K(t)=K_0 \,t^{-\alpha}, \qquad 0<\alpha<1,
\label{FLE}
\end{eqnarray}
in which case  the FLE describes  subdiffusion~\cite{Porra,Hansen}.
The term $F(t)$ in (\ref{GLE}) is a zero-centered stationary  noise, which is related to the kernel by the fluctuation-dissipation theorem
$\langle F(t)F(t')\rangle=k_BT K(t-t')$ where
$T$ is temperature.
If an external force is also applied (not considered here), it is assumed 
not to modify the kernel.
Equation (\ref{GLE}) with kernel (\ref{FLE}) is called "fractional" because if $A=\dot a$ then
the nonlocal term in (\ref{GLE}) is proportional to the Caputo fractional derivative 
$\frac{d^\alpha}{dt^\alpha} \,a(t)=
\frac{1}{\Gamma(1-\alpha)}\,\int_0^t (t-\tau)^{-\alpha}\,\dot a(\tau)\, d\tau$.  
Although writing the FLE in terms of the fractional derivative may be insightful~\cite{Lutz}, 
the equation can be, and often is,
exploited  with no tools specific to fractional calculus.

Among other types of generalized Langevin equations, the FLE is distinguished by the diverging integral
$\gamma=\int_0^\infty K(t)\, dt$, which in other cases gives the friction constant $\gamma$
in  the  Markovian approximation.
Therefore, there is no Markovian approximation for the FLE.
A physical consequence of this mathematical property, as was noticed above, is 
anomalous diffusion:
If $A$ stands for the velocity $\dot x$ of a Brownian object, then
the FLE predicts subdiffusion, i.e. the mean-square displacement $\langle x^2(t)\rangle$ increasing sublinearly, namely as $t^{\alpha}$~\cite{Porra,Hansen}. Another peculiarity is that stochastic processes governed by the FLE exhibit nontrivial ergodic properties~\cite{Deng}.

As a matter of fact, the aforementioned unique features of the FLE still hold if the power-law dependence (\ref{FLE}) takes place not for the entire time domain, but only asymptotically at long times, 
\begin{eqnarray}
K(t)\sim t^{-\alpha},\qquad 0<\alpha<1,
\qquad \mbox {as}\quad t\to\infty.
\label{FLE2}
\end{eqnarray}
One may call equations
with  kernels (\ref{FLE2})   asymptotically fractional, but we prefer to keep to the established term FLE for such equations 
as well, even though in that case the nonlocal term in (\ref{GLE}) may lose the meaning of a fractional derivative.

The FLE was first introduced on a phenomenological basis to describe 
anomalous diffusion in geometrically disordered 
and fractally organized 
systems like percolation clusters~\cite{Nakanishi}.
The existence of a dynamical theory giving rise to the FLE is far from obvious and was explicitly doubted~\cite{Mazo}. 
Later studies, however, suggested that
the origin of the FLE in many phenomena, particularly protein  conformational transitions, may be purely dynamical~\cite{Kneller,Xie}.
Is there a specific feature of inherent dynamics in complex systems characteristic of  subdiffusion and emergence of the FLE?


A standard model to formally derive
a generalized Langevin equation with a memory kernel of {\it any} assigned form is that  of 
a Brownian particle linearly 
coupled to the bath 
comprised of independent harmonic 
oscillators~\cite{Zwanzig,Weiss}.
In such a setting, the memory kernel can be  expressed as a Fourier transform of a certain function $C(\omega)$, describing 
spectral properties of the bath
and bath's coupling to a system (see below).
Assuming that 
\begin{eqnarray}
C(\omega)=C_0\,\omega^{\beta}, \qquad -1<\beta<0,
\label{condition_0}
\end{eqnarray}
 one recovers the fractional kernel (\ref{FLE}) with the exponent $\alpha=1+\beta$~\cite{Kup}.
Other approaches to derive the FLE  were considered in~\cite{Weiss,Lizana,Taloni}.


In this paper we show that with additional dissipative forces acting on the bath oscillators 
the aforementioned 
standard model leads to the FLE
under a much less restrictive condition. It still
has the form (\ref{condition_0}), but for  a larger range of $\beta$, namely $ -1<\beta<1$. For the 
the fractional exponent our model  predicts $\alpha=(1+\beta)/2$ instead of $\alpha=1+\beta$ for the standard model.
More importantly, the power-law  dependence of the spectral function $C(\omega)$ is
required not for the entire frequency range (as in the standard model), but only asymptotically 
in the limit $\omega\to 0$, see Eq. (\ref{condition}) below.  We argue that a rather relaxed character of this condition may explain the omnipresence of fractional stochastic dynamics in a wide range of complex systems.

We start with an observation that 
in many systems exhibiting fractional stochastic dynamics 
the thermal bath involves two groups of dynamical variables evolving on  well-separated timescales. 
The separation of timescales for a system of interest on the one hand, and for the bath variables on the other hand is, of course, a common feature in many statistical mechanical models. In contrast, a specific assumption of the presented  model is that the separation of timescales takes place for the bath alone. For brevity, we shall refer to the parts of the bath 
comprising slow and fast variables as
the slow and fast baths, respectively, assuming that both baths have the same temperature. 
For a macromolecule in a solvent~\cite{Xie,Kneller},
the slow bath refers to slowly evolving degrees of freedom of the macromolecule itself, while 
the fast bath is  comprises molecules of the solvent as well as fast degrees of of freedom of the macromolecule. 
Schematically, the model with such a
double hierarchy of the bath is depicted in Fig. 1.

We shall assume that: (1) the slow bath has no characteristic timescale except the lower cut-off value  $t_0$ (corresponding to the highest-frequency mode~\cite{Weiss}), and 
(2) a characteristic time $t_1$ of the fast bath does exist and is  much shorter than $t_0$, $t_1\ll t_0$.
It is not required
that the two baths are
independent. On the contrary, the coupling of the slow bath variables to the fast bath will be shown to be 
essential for the emergence of the fractional dynamics. 
On the other hand, we shall assume that
the influence of the slow bath on dynamics of the fast one is negligible. 
Under these  assumptions it is
natural to describe the dynamics of slow and fast baths in different manners. Namely, the dynamics of 
slow bath variables will be taken into account explicitly, while
effects  of the fast bath will be modeled implicitly
by adding Markovian Langevin forces, 
i.e. in the same way as in the Rouse model used in polymer physics.


We describe the slow bath as a set of $N$ independent oscillators, linearly coupled to a system of interest, which we shall call the "particle".  The Hamiltonian of the particle and the slow bath we choose to be of  the Caldeira-Leggett form~\cite{Zwanzig,Weiss}
\begin{eqnarray}
H&=&\frac{1}{2M}\,P^2+H_b,\label{H}\\
H_b&=&\sum_{i=1}^N \left\{
\frac{1}{2}\,p_i^2+\frac{\omega_i^2}{2}
\left(
q_i-\frac{\gamma_i}{\omega_i^2}\,Q
\right)^2
\right\}.
\label{H_b}
\end{eqnarray}
Here $(Q, P)$
and $\{q_i, p_i\}$
are the coordinates and momenta of the particle and 
slow bath oscillators, respectively, $\omega_j$  oscillators frequencies, $\gamma_i$ coupling constants, $M$ the mass of the particle, and the masses of all oscillators are set equal to one. A physically plausible interpretation of this  Hamiltonian 
suggests that
the independent oscillators  represents collective normal modes of the slow bath rather than the bath's individual constituents.

The equations of motion corresponding to the above Hamiltonian read
\begin{eqnarray}
\dot P&=&\sum_{i=1}^N \gamma_i \left(
q_i-\frac{\gamma_i}{\omega_i^2}\, Q\right),
\label{eom_system}\\
\dot p_i&=&-\omega_i^2\,q_i+\gamma_i\, Q.
\label{eom_bath}
\end{eqnarray}
These equations are identical to those of the standard oscillator bath model~\cite{Zwanzig,Weiss}  and yet do not take into account the presence  of the fast bath. The effects of the latter we describe by adding time-local Langevin forces 
\begin{eqnarray}
f_i=-2\lambda\, p_i+\xi_i,
\label{L_force}
\end{eqnarray}
acting on $i$-th oscillator of the slow bath.
Here the white noise forces $\xi_i(t)$ are zero-centered and related to the 
damping coefficient $\lambda$ (assumed to be the same for all oscillators)
by the fluctuation-dissipation relations
\begin{eqnarray}
\langle \xi_i(t)\,\xi_{i'}(t')
\rangle=
4 \lambda\,k_BT\,\delta(t-t')\,\delta_{ii'}.
\label{fdt}
\end{eqnarray}
Here the Kronecker symbol $\delta_{ii'}$ implies that stochastic forces $\xi_i$ acting on different oscillators are uncorrelated.

\begin{figure}[t]
\includegraphics[height=3.8cm]{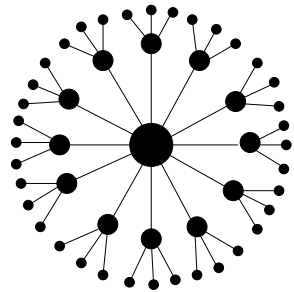}
\caption{The idealized scheme of the model:
A systems of interest (the large central circle) is directly coupled to 
oscillators of the slow bath (medium-size inner circles), which
in turn are  coupled to particles of the fast bath  (small outer circles). The latter are taken into 
account implicitly 
\`a la Langevin.
}
\end{figure}

With the Langevin forces added, the equations of motions for the slow bath oscillators (\ref{eom_bath}) take the form
\begin{eqnarray}
\dot p_i=-\omega_i^2q_i+\gamma_i\, Q -2\lambda\,p_i+\xi_i.
\end{eqnarray}
Replacing momenta by  velocities, $p_i=\dot q_i$, one can re-write this as equations for $q_i$,
\begin{eqnarray}
\ddot q_i+2\lambda\,\dot q_i+\omega_i^2 q_i=\gamma_i\, Q +\xi_i.
\label{eom_bath2}
\end{eqnarray}
In our model, these are  equations of motion of the slow bath oscillators coupled to the particle and  to the fast bath.

We shall assume that
an equation of motion for the particle still has the form (\ref{eom_system}), i.e. 
the particle is directly coupled to the slow bath only (as depicted in Fig. 1). This assumption is not essential:  a  
coupling of the particle to the fast bath
can be easily taken into account as well, but
it would have a trivial effect 
of an additional delta-function contribution
in the final expression for the memory kernel,
which does not affect the asymptotic long-time properties of the model.

Solving Eqs. (\ref{eom_bath2}), for instance by the method of Laplace transforms,  one finds
\begin{eqnarray}
\!q_i(t)\!=\!q_i^0(t)\!+\!\int_0^t\! b_i(t\!-\!\tau)\,\Big(\gamma_i\,Q(\tau)+
\xi_i(\tau)\Big)\,d\tau.
\label{q_solution}
\end{eqnarray}
Here  $q_i^0(t)$ is a solution of the corresponding homogeneous equation
$\ddot q_i+2\lambda\,\dot q_i+\omega_i^2 q_i=0$,
which we write as
\begin{eqnarray}
q_i^0(t)&=&a_i(t)\,q_i(0)+b_i(t)\,\dot q_i(0),
\end{eqnarray}
where the functions $a_i(t)$ and $b_i(t)$ are 
\begin{eqnarray}
&&a_i(t)=e^{-\lambda t}\,\left(
\cosh\Omega_i t+\frac{\lambda}{\Omega_i}\,
\sinh\Omega_i t
\right),
\label{a}
\\
&&b_i(t)=\frac{e^{-\lambda t}}{\Omega_i}\,\sinh\Omega_i t.
\label{b}
\end{eqnarray}
In these expressions,  the frequency-like parameter 
\begin{eqnarray}
\Omega_i=\sqrt{\lambda^2-\omega_j^2}
\label{Omega}
\end{eqnarray}
is real for overdamped ($\omega_i<\lambda$) and imaginary for  underdamped 
($\omega_i>\lambda$) oscillators, yet
in both cases the functions
 $a_i(t)$ and $b_i(t)$ are real-valued.


Expression (\ref{q_solution}) for $q_i(t)$ is intended to be substituted into the equation of motion (\ref{eom_system}) in order to bring the latter
into a Langevin form. To this end, the standard trick is to integrate
the term with $Q$ in  (\ref{q_solution}) by parts.
Taking into account that an  anti-derivative of $b_i(t)$ equals
\begin{eqnarray}
B_i(t)=\int_0^t\!b_i(\tau)\, d\tau=\frac{1}{\omega_j^2}\Big(
1-a_i(t)\Big),
\end{eqnarray}
one brings expression (\ref{q_solution}) 
into  the form
\begin{eqnarray}
q_i(t)\!\!&=&\!\!q_i^0(t)
\!-\!\frac{\gamma_i}{\omega_i^2}\int_0^t\!\! a_i(t\!-\!\tau) \dot Q(\tau)\,d\tau\!+\!\frac{\gamma_i}{\omega_i^2}\,Q(t)
\nonumber\\
&-&\!\!\frac{\gamma_i}{\omega_i^2}\, a_i(t)\,Q(0)+
\int_0^t \!\! b_i(t\!-\!\tau)\,\xi_i(\tau)\, d\tau.
\end{eqnarray}
Substitution of this expression into the particle's equation of motion (\ref{eom_system})
yields 
the generalized Langevin equation
\begin{eqnarray}
\dot P(t)=-\int_0^t K(t-\tau)\,P(\tau)\, d\tau+F(t),
\end{eqnarray}
with the memory kernel
\begin{eqnarray}
K(t)=\frac{1}{M} \sum_{i=1}^N \left(
\frac{\gamma_i}{\omega_i}
\right)^2 a_i(t)
\label{kernel1}
\end{eqnarray}
and the noise force
\begin{eqnarray}
F(t)\!=\!\sum_{i=1}^N \gamma_i\left\{ 
q_i^0(t)\!-\!
\frac{\gamma_i}{\omega_i^2}\,a_i(t) Q(0)
\!+\!\!\int_0^t\!\! b_i(t\!-\!\tau)\xi_i(\tau)d\tau
\right\}.\nonumber
\end{eqnarray}
If there is no fast bath, then $(\lambda,\, \xi_i)\to 0$, and the above expressions coincide with that for the standard oscillator bath model~\cite{Zwanzig,Weiss}.    Suppose the initial state of the slow bath is described by a canonical ensemble with the Hamiltonian $H_b$ given by (\ref{H_b}), then 
one can show that
the noise $F(t)$ is zero-centered, $\langle F(t)\rangle=0$, and the fluctuation-dissipation relation between $F(t)$ and $K(t)$ can be readily
established.

Next, we make a usual assumption that in the limit $N\to\infty$
the spectrum of the slow bath becomes continuous, $\{\omega_i\}\to \omega$, and 
the sums over $i$ can be replaced by integrals,  $\sum_i\to\int_0^\infty d\omega\,g(\omega)(\cdots)$, where $g(\omega)$ is the density of the slow bath states~\cite{Zwanzig,Weiss,Mazur}.
Replacing the coupling constants and functions $a_i(t)$ of individual oscillators by smooth functions of frequency,
\begin{eqnarray}
\gamma_i\to \gamma(\omega),\qquad
a_i(t)\to a(t,\omega),
\end{eqnarray}
the memory kernel (\ref{kernel1}) 
can be written in the form 
\begin{eqnarray}
K(t)=\frac{1}{M}\,\int_0^\infty \!\!d\omega\,g(\omega)\,\frac{\gamma(\omega)^2}{\omega^2}\, a(t,\omega).
\label{kernel2}
\end{eqnarray}
We re-write this more compactly  as
\begin{eqnarray}
K(t)=\int_0^\infty \!\!d\omega\, C(\omega)\, a(t, \omega)
\label{kernel3}
\end{eqnarray}
defining the function 
\begin{eqnarray}
C(\omega)=
\frac{1}{M}\,g(\omega)\,\frac{\gamma(\omega)^2}{\omega^2}
\label{C}
\end{eqnarray}
which accumulates the spectral properties of the slow bath.

As follows from (\ref{a}) and (\ref{kernel3}), the Laplace transform of the kernel $\tilde K(s)=\int_0^\infty e^{-st} K(t)\, dt$ reads
\begin{eqnarray}
\tilde K(s)=\int_0^\infty \!\!d\omega\,C(\omega)\, 
\frac{s+2\lambda}{s^2+2\lambda s+\omega^2}.
\label{kernel4}
\end{eqnarray}
According to a Tauberian theorem~\cite{Feller}, 
the long-time behavior of the kernel $K(t)$
is determined by the behavior of its Laplace transform $\tilde K(s)$
at small $s$ as follows:
\begin{eqnarray}
\!\!\!\!\!\!
\tilde K(s)\sim s^{-\gamma}\,\,\mbox{as}\,\,\, s\to 0
\,\,\,\,
\Rightarrow
\,\,\,\,
K(t)\sim t^{\gamma-1}\,\,\, \mbox{as}\,\,\, t\to\infty.
\label{Tauberian}
\end{eqnarray}
In order to find an asymptotic form of $\tilde K(s)$ for small s, one can neglect
$s$ in the numerator and $s^2$ in the denominator of the integrand of (\ref{kernel4}). Then it can be written as 
\begin{eqnarray}
\tilde K(s)\!=\!\pi\sqrt{\frac{2\,\lambda}{s}}
\int_0^\infty\!\!\! d\omega\,C(\omega)\, L\left(\omega,\sqrt{2\lambda s}\right), 
\label{kernel5}
\end{eqnarray}
where $L(\omega,\Gamma)$ is the zero-centered Lorentzian (or Cauchy) distribution
\begin{eqnarray}
L(\omega,\Gamma)=\frac{1}{\pi}\,\frac{\Gamma}{\omega^2+\Gamma^2}.
\end{eqnarray}
Since $L\left(\omega,\Gamma\right)$ tends to the delta-function $\delta(\omega)$ as $\Gamma\to 0$,  it is intuitively clear that, unless the function  
$C(\omega)$ increases too fast, 
only the low-frequency behavior of $C(\omega)$ is matter 
for the asymptotic evaluation of integral 
(\ref{kernel5}) as $s\to 0$. 

Consider the spectral function of the asymptotic form
\begin{eqnarray}
C(\omega)\sim \omega^\beta, \quad -1<\beta<1, \qquad \mbox{as}\,\,\,\omega\to 0.
\label{condition}
\end{eqnarray}
The only condition we impose on $C(\omega)$  at
nonsmall frequencies is the convergence of integral
(\ref{kernel5}). For such $C(\omega)$ the integrand $D(\omega)=C(\omega) L(\omega, \sqrt{2\lambda s})$  of (\ref{kernel5}) behaves differently for  the two subintervals of $\beta$:

(a) $-1<\beta<0$:  $D(\omega)$
has an integrable singularity at the lower integration limit $\omega=0$;

(b) $0<\beta<1$: $D(\omega)$
has a maximum at $\omega_m>0$.

\noindent These two types of behavior of $D(\omega)$ actually become qualitatively similar for small $s$ because for case (b) the position of the maximum  $\omega_m$ approaches  zero 
and the maximum value diverges
as $s\to 0$, see Fig. 2.

One can show that in both cases the main contribution to the integral (\ref{kernel5}) comes from a neighborhood of $\omega=0$ and has the asymptotic form
\begin{eqnarray}
\tilde K(s)\sim s^{-\gamma}, 
\quad 
\gamma=\frac{1}{2}(1-\beta)
\label{kernel6}
\end{eqnarray}
as $s\to 0$. If the power law  (\ref{condition}) 
with $0<\beta<1$, case (b),
holds for the entire frequency range, than the result (\ref{kernel6}) 
can be verified directly substituting 
$C(\omega)=C_0\,\omega^\beta$ into (\ref{kernel5}).

\begin{figure}[t]
\includegraphics[height=5.1cm]{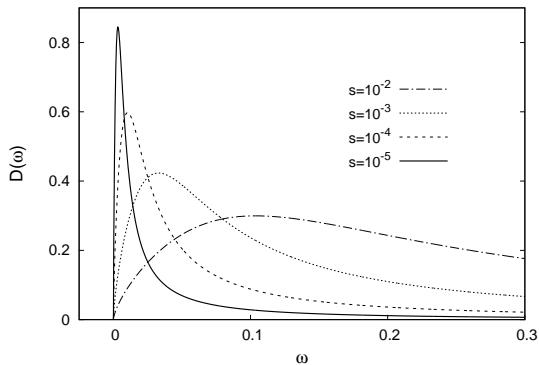}
\caption{The integrand function 
$D(\omega)=C(\omega)\,L(\omega, \sqrt{2\lambda s})$  of the kernel (\ref{kernel5}) for  $C(\omega)=\omega^{\beta}$ with $\beta=0.7$ and several values of the Laplace variable $s$ (arb. units).}
\end{figure}

In a far more general and interesting case when the power dependence  of $C(\omega)$ holds only asymptotically for small $\omega$,
as condition (\ref{condition}) requires, the result (\ref{kernel6}) can be justified,
for both cases 
(a) and (b), as follows. Let us 
split the integral (\ref{kernel5})
into two parts, with the integration ranges $(0,\epsilon)$ and $(\epsilon, \infty)$.
For large $\omega$ the factor $L(\omega,
\sqrt{2\lambda s})/\sqrt{s}$ in (\ref{kernel5})
depends  on $s$ only weakly.
It is therefore natural to assume that the second part, involving integration over  $(\epsilon, \infty)$, gives a contribution which is bounded for $s\to 0$.
Then for an arbitrary small but
fixed $\epsilon$ and for $s\to 0$ the kernel
$\tilde K(s)$ is determined by the first part,
diverging for small $s$,
\begin{gather}
\!\!\!\!\!\!\!\!\!\!\!\!\!\!\!\!\!
\!\!\!\!\!\!\!\!\!\!\!\!\!\!\!\!\!
\tilde K(s)\approx
\pi\sqrt{\frac{2\lambda}{s}}
\int_0^\epsilon C(\omega)\, L\left(\omega,\sqrt{2\lambda s}\right)\,d\omega\nonumber\\
\qquad\propto{}_2F_1\left(1,\frac{1+\beta}{2};\frac{3+\beta}{2}; -\frac{\epsilon^2}{2\lambda s}\right)\, \frac{\epsilon^{1+\beta}}{(1+\beta)\,s}.
\label{hyper}
\end{gather}
Taking into account asymptotic properties of the Gauss hypergeometric function  ${}_2F_1(a,b;c;z)$ at large $z$~\cite{Temme},
one recovers from (\ref{hyper})
the power-law asymptotic behavior (\ref{kernel6}).

Finally,  using the Tauberian 
theorem (\ref{Tauberian}), one finds
from (\ref{kernel6}) that in the time domain
the kernel has a fractional asymptotic form,
\begin{eqnarray}
K(t)\sim t^{-\alpha}, 
\quad 
\alpha=1-\gamma=\frac{1}{2}(1+\beta).
\label{kernel7}
\end{eqnarray}
Since we assume $-1<\beta<1$,
then 
$0<\alpha<1$.

The above asymptotic arguments can be illustrated with specific spectral functions
defined for the entire frequency range.
As an example consider 
\begin{eqnarray}
C(\omega)=\frac{C_0}{1+\tau^2\,\omega^2}.
\label{example1}
\end{eqnarray}
In the standard oscillator bath model, the memory kernel is a Fourier transform of $C(\omega)$,
$K(t)=\int_0^\infty \!d\omega \,C(\omega)\, \cos(\omega t)$, and for the spectral function (\ref{example1}) takes the exponential form $K(t)=K_0\,e^{-|t|/\tau}$~\cite{Weiss}. 
Instead, in the present model 
the corresponding kernel is fractional
and its dependence on $\tau$ disappears:
since $C(\omega)\sim \omega^0$ as $\omega\to 0$,  then $\beta=0$, and Eq. (\ref{kernel7}) predicts
$K(t)\sim t^{-1/2}$.
The exact evaluation of the kernel by substituting
$C(\omega)$ of the form (\ref{example1}) into (\ref{kernel5}) 
confirms this result.

Another example is the spectral function $C(\omega)=C_0\,\omega^\beta\, e^{-\tau \omega}$ with $-1<\beta<1$.
Again, the direct evaluation of the kernel using (\ref{kernel5}) confirms the asymptotic kernel's behavior  (\ref{kernel7}). 

In the above examples, the emergence of  fractional 
kernel (\ref{kernel7}) 
depends neither on $\tau$, nor on $\lambda$
(the parameters 
characterizing the slow and fast baths, respectively), but only on low-frequency
spectral properties of the slow bath. This illustrates
the robustness  of fractional dynamics in the presented model.

In conclusion, the presented model shows that the fractional Langevin  dynamics may emerge 
under much broader conditions than some earlier models suggested.
The conditions concern the bath's asymptotic  low-frequency properties only.
Although modeling of the slow bath as a set of independent oscillators may appear on first glance unrealistic, it is in fact physically well-motivated and relevant for systems like lattices and networks,  whose Hamiltonian can be diagonalized, exactly or approximately,  into a sum of contributions from independent collective normal modes. As a simple illustration one may  point to 
the familiar problem of 
a tagged particle, or an isotope, in an otherwise uniform one-dimensional harmonic lattice immersed in a fluid,
The problem can be mapped into 
the model discussed in this paper
with the chain's normal modes serving 
as independent oscillators of the 
slow bath, and the fluid as a fast Langevin bath. 
For the linear harmonic chain  the explicit expressions for 
the normal mode $g(\omega)$ and coupling $\gamma(\omega)$ distributions are well-known~\cite{Mazur}. Then
for the spectral function (\ref{C})
one finds the asymptotic dependence $C(\omega)\sim\omega^0$ as $\omega\to 0$. Accordingly $\beta=0$,
and for the tagged particle's momentum our model predicts the FLE with the fractional exponent $\alpha=(1+\beta)/2=1/2$.  
For an over-damped chain this is a 
well-known result~\cite{Lizana}, but here we get it for an arbitrary level of damping.



\end{document}